\documentstyle[12pt,epsfig]{article}
\newcommand{\be}{\begin{equation}}
\newcommand{\ee}{\end{equation}}
\newcommand{\bea}{\begin{eqnarray}}
\newcommand{\eea}{\end{eqnarray}}
\newcommand{\ba}{\begin{array}}

\newcommand{\ea}{\end{array}}

\def\bbox{{\,\lower0.9pt\vbox{\hrule \hbox{\vrule height 0.2 cm
\hskip 0.2 cm \vrule height 0.2 cm}\hrule}\,}}
\newcommand{\dsl}{\pa \kern-0.5em /}

\newcommand{\nn}{\nonumber \\}
\newtheorem{theorem}{Theorem}[section]
\newtheorem{prop}[theorem]{Proposition}

\newtheorem{define}[theorem]{Definition}

 \newcommand{\beq}{\begin{equation}} 
\newcommand{\eeq}{\end{equation}} \newcommand{\beqn}{\begin{eqnarray}} 
\newcommand{\eeqn}{\end{eqnarray}}

\newcommand{\bI}{{I}}  
  
\newcommand{\bmu}{{m}}  
\newcommand{\bnu}{{n}} 
\newcommand{\bp}{{p}} 
\newcommand{\bq}{{q}}    

\begin{document}
\baselineskip 18pt
\begin{titlepage}
\begin{flushright}
Duke-CGTP-00-06\\
KIAS-P00022\\
hep-th/0005275\\
\end{flushright}

\vfill

\begin{center}
\baselineskip=16pt
{\Large\bf Counting Yang-Mills Dyons with Index Theorems} \\
\vskip 10.mm
{ ~Mark Stern$^{*,1}$ and ~Piljin Yi$^{+,2}$ } \\
\vskip 0.5cm
{\small\it 
$^*$ Department of Mathematics, Duke University, Durham, NC 27706, USA
}\\
\vspace{12pt}
{\small\it
$^+$
School of Physics, Korea Institute for Advanced Study\\
207-43, Cheongryangri-Dong, Dongdaemun-Gu, Seoul 130-012, Korea
}\\
\vspace{6pt}
\vfill
\end{center}
\par
\begin{center}
{\bf ABSTRACT}
\end{center}
\begin{quote}
We count the supersymmetric bound states of many distinct BPS monopoles in 
$N=4$ Yang-Mills theories and in pure $N=2$ Yang-Mills theories. The
novelty here is that we work in generic Coulombic vacua where more than
one adjoint Higgs fields are turned on. The number of purely magnetic bound 
states is again found to be consistent with the electromagnetic duality of 
the $N=4$ $SU(n)$ theory, as expected. We also
count dyons of generic electric charges, which correspond to 1/4 BPS 
dyons in $N=4$ theories and 1/2 BPS dyons in $N=2$ theories. Surprisingly,
the degeneracy of dyons is typically much larger than would be accounted 
for by a single supermultiplet of appropriate angular momentum, implying
many supermutiplets of the same charge and the same mass.
\vfill
\vskip 5mm
\hrule width 5.cm
\vskip 5mm
{\small
\noindent $^1$ E-mail: stern@math.duke.edu, \\
\noindent $^2$ E-mail: piljin@kias.re.kr\\
}
\end{quote}
\end{titlepage}
\setcounter{equation}{0}

\section{Introduction}

The purpose of this paper is to count BPS monopoles and dyons in 
supersymmetric Yang-Mills theories. Such computations have been performed
by many authors using moduli space dynamics of monopoles
\cite{sen,sethi,GH,lwy,gibbons}. Recently \cite{tong,blly,n2}, 
however, it was realized that the monopole dynamics in a generic
vacuum is qualitatively different from the old moduli space dynamics of
Manton \cite{manton} employed in most such endeavors, where the low energy
dynamics of monopoles were considered when only one adjoint Higgs field
is turned on, while supersymmetric Yang-Mills field theories 
come with 2 or 6 such scalars. This restriction
disallows static interaction between monopoles \cite{ejw}, so that all 
interaction comes from nontrivial coefficients of kinetic terms.
In a generic vacuum with more than one adjoint Higgs turned on, 
monopoles of the same type still have no static force among them,
but dynamics of monopoles of distinct type could have a static potential.
In this article, we solve various index problems as a first step towards
counting all BPS states.

In the old monopole dynamics, the Lagrangian one finds is a pure 
sigma model with the moduli space as the target manifold.
Classically, one solves for geodesic trajectories to find classical 
orbits of monopoles. Quantum mechanically, the Hamiltonian is a square
of a supercharge which can be regarded as a Dirac operator 
acting either on the spinor bundle or on the Clifford bundle over the moduli
space,
\begin{equation}
Q\sim -i\gamma^m\nabla_m.
\end{equation}
Supersymmetric bound states, for example, would be found as
normalizable spinors or forms on the moduli space that are 
also zero modes of this Dirac operator.

The new supersymmetric low energy dynamics is obtained by augmenting
old moduli space dynamics with a set of supersymmetric potential terms,
and was written explicitly in Ref.\cite{blly,n2}. 
In principle, the question of
BPS states must be reconsidered in the new dynamics. (One example of
states that cannot be probed in the old formalism is the now well-known
1/4 BPS states \cite{bergman,ly,hashimoto}.)
In this new setting, the supercharges of the low energy dynamics will 
again be interpreted as Dirac operators on the moduli space, which is now
twisted by some triholomorphic Killing vector field, say $K$, 
\begin{equation}
Q\sim \gamma^m(-i\nabla_m -K_m).
\end{equation}
The corresponding bosonic potential is precisely half the squared norm of 
$K$ \cite{gaume,tong,blly}. 
We count the number of normalizable states annihilated by a Dirac operator, 
weighted 
by $\pm 1$ for the chiral and the antichiral states respectively. 
The resulting integer is the index of the Dirac operator. In fact this 
index would be infinite or ill defined without further restriction of the 
domain of the Dirac operator. We restrict the problem to each charge 
eigensector before counting the zero modes of the Dirac operator. Thus,
effectively, we will be computing an equivariant index with $L^2$ condition. 
This gives information on the existence and the degeneracy of dyonic 
bound states for each electric charge sector.

Understanding monopole dynamics in generic vacua is particularly significant 
in the context of $N=2$ Yang-Mills theories, because many (1/2) BPS dyons
exist only when both adjoint Higgs are turned on. 
By ignoring the potential term in the low energy dynamics, one would in effect 
be searching for a bound state in the vacua where the state cannot exist as 
a supersymmetric one-particle state. In the language of Seiberg-Witten 
\cite{sw}, one would be looking for it on one side of a marginal stability 
domain wall in the vacuum moduli space, while the bound state in question 
exists only on the other side of the domain wall. For example,
most dyons that become massless at special hypersurfaces in the 
Seiberg-Witten moduli space, are of this type \cite{su3,lerche,hollo} and 
cannot be probed by old moduli space dynamics.

In Section 2, we review the supersymmetric quantum mechanics
with potential for the case of 4 and 8 supercharges. We isolate
various involutions, with respect to which the index is defined.
Section 3 introduces the explicit form of the moduli space metric 
that governs the dynamics of BPS monopoles. The only known moduli space
for arbitrarily many monopoles is the case of all distinct monopoles,
and this is the case for which we will compute the index explicitly. 
Section 4 recalls a recent explicit computation of 
two-monopole bound states in $SU(3)$ theories and presents the 
resulting value of the indices. In section 5, we finally delve into 
the computation of the index by using a Fredholm deformation of the 
Dirac operator in question. The computation reduces to that of a  
superharmonic oscillator in 4 dimensions, whose index is computed
explicitly. Section 6 translates the
results to degeneracies of various dyonic and purely magnetic
bound states and checks its consistency with anticipated nonperturbative
physics. We close with a summary.

\section{Supersymmetric Sigma Model with Potential}

In this section, we briefly review the supersymmetric sigma-model
quantum mechanics with potentials. These quantum mechanics have been 
introduced as the low energy dynamics of monopoles in pure $N=2$ 
Yang-Mills field theory and in $N=4$ Yang-Mills field theory 
\cite{blly,n2}. They also appeared in other systems such as the 
dynamics of instanton solitons \cite{nci}.

\subsection{Quantum Mechanics with 4 Real SUSY}

The SUSY dynamics we consider is a sigma model with potential, whose
Lagrangian is written compactly as 
\begin{equation}
{\cal L}={1\over 2} \biggl( g_{\bmu\bnu} \dot{z}^\bmu \dot{ z}^\bnu +
ig_{\bmu\bnu} \lambda^\bmu D_t \lambda^\bnu
 - g^{\bmu\bnu} G_\bmu G_\bnu - i\nabla_\bmu G_\bnu  
\lambda^\bmu \lambda^\bnu  \biggr),
\label{action}
\end{equation}
where $D_t\lambda^\bmu=\dot \lambda^\bmu +\Gamma^\bmu_{\bnu\bp}\dot z^\bnu
\lambda^\bp$. The target manifold must be hyperK\"ahler, which means
that there are three covariantly constant complex structures $J^{(s)}$ 
satisfying quaternionic algebra,
\begin{equation}
J^{(s)}J^{(t)}=-\delta^{st}+\epsilon^{stu}J^{(u)},
\end{equation}
and the Killing vector field $G$ should be triholomorphic.
\begin{eqnarray} 
{\cal L}_{G} g=0, \qquad {\cal L}_{G} J^{(s)}=0.
\end{eqnarray}

Introducing vielbein $e^{E}_\bmu$ and defining $\lambda^{E}=\lambda^\bmu 
e_\bmu^{E}$ which commute with all bosonic variables, the canonical
commutators are
\begin{eqnarray}
[z^\bmu,p_\bnu]&=&i\delta^\bmu_\bnu  , \nn
\{\lambda^{E},\lambda^{F}\}&=&\delta^{EF} .
\end{eqnarray}
We can realize this algebra on spinors on the moduli space by
letting $\lambda^{E}=\gamma^{E}/{\sqrt 2}$, where $\gamma^{E}$ 
are gamma matrices.
(Since the moduli space is hyperK\"ahler an equivalent quantization
is obtained using holomorphic differential forms.)
The supercovariant momentum operator, defined by
\begin{equation}
\pi_\bmu=p_\bmu-{i\over 4}\omega_{\bmu EF}[\lambda^{E},\lambda^{F}] ,
\end{equation}
where $\omega_{\bmu\, E}^{\, \, F}$ is the spin connection, then becomes
the covariant derivative acting on spinors $\pi_\bmu=-i\nabla_\bmu$. Note
that
\begin{eqnarray}
{[}\pi_\bmu,\lambda^\bnu{]}&=&i\Gamma^\bnu_{\bmu\bp}\lambda^\bp ,\nn
{[}\pi_\bmu,\pi_\bnu{]}&=&
-{1\over 2}R_{\bmu\bnu\bp\bq}\lambda^\bp\lambda^\bq .
\end{eqnarray}

The supersymmetry charges take the form
\begin{eqnarray}
Q&=&\lambda^\bmu(\pi_\bmu-G_\bmu) ,\nn
Q^{(s)}&=&\lambda^\bmu {J^{(s)\bnu}_{\,\, \bmu}}(\pi_\bnu-G_\bnu) ,
\end{eqnarray}
which obey
\begin{eqnarray}
\label{algebra}
\{Q,Q\}&=&2({\cal H}-{\cal Z})  , \nn
\{Q^{(s)},Q^{(t)}\}&=&2\,\delta_{st}({\cal H}-{\cal Z}) , \nn
\{Q,Q^{(s)}\}&=&0 .
\end{eqnarray}
Here the Hamiltonian ${\cal H}$ and the central charge ${\cal Z}$ are
given by
\begin{eqnarray}
&&{\cal H}=
{1\over 2} \biggl( {1\over \sqrt{g}}\pi_\bmu \sqrt{g }g^{\bmu\bnu}\pi_\bnu
+ G_\bmu G^\bmu  + i\nabla_\bmu G_\bnu \lambda^\bmu\lambda^\bnu \biggr),\\
&& {\cal Z}= G^\bmu \pi_\bmu -{i\over 2}  
(\nabla_\bmu G_\bnu)\lambda^\bmu\lambda^\bnu.
\label{hamiltonian}
\end{eqnarray}
Note that the operator $i{\cal Z}$ is  the Lie derivative ${\cal L}_G$
acting on spinors (see {\it e.g.}, \cite{ggpt})
\begin{equation}
{\cal L}_G\equiv G^m\nabla_m+
{1\over 8}\nabla_\bmu G_\bnu[\gamma^\bmu,\gamma^\bnu] .
\end{equation}
The SUSY quantum mechanics comes with a natural $Z_2$ grading defined
by the operator,
\begin{equation}
\tau_2=\prod 2^{1/2}\lambda^E =\prod \gamma^E,
\end{equation}
which anticommutes with the Dirac operator,
\begin{equation}
D=\sqrt{2}\,Q=\gamma^\bmu(-i\nabla_\bmu-G_\bmu) .
\end{equation}
This pair defines the Witten index that counts the difference between
the number of bosonic states and the number of fermionic states annihilated
by the supercharge. In fact, the index is defined in each superselection 
sector with fixed ${\cal Z}$, and effectively counts the difference in 
the numbers of BPS states of given central charges. The index will be
denoted collectively by ${\cal I}_2$. See Section 5 for detailed computation
of ${\cal I}_2$.

\subsection{Quantum Mechanics with 4 Complex SUSY}

When the number of supercharges and the number of fermions double,
we obtain the following form of sigma model with potential,
\begin{eqnarray}  
{\cal L}&=&{1\over 2} \biggl( g_{\bmu\bnu} \dot{z}^\bmu \dot{ z}^\bnu +  
ig_{\bmu\bnu} \bar\psi^\bmu \gamma^0 D_t \psi^\bnu + {1\over 6}  
R_{\bmu\bnu\bp\bq}\bar\psi^\bmu \psi^\bp \bar\psi^\bnu \psi^\bq  \nonumber \\
&&- g_{\bmu\bnu} G_{\bI}^\bmu   
G_{\bI}^\bnu  
-i \nabla_\bmu G_{\bI \bnu} \,\, \bar\psi^\bmu  
(\Omega^\bI \psi)^\bnu  
\biggr),  
\label{susyaction}  
\end{eqnarray}  
where $\psi^\bmu$ is a two component Majorana spinor, $\gamma^0=\sigma_2, 
\gamma^1=i\sigma_1, \gamma^2=-i\sigma_3$, $\bar\psi=\psi^T\gamma^0$. 
The operator $\Omega_\bI$'s are defined respectively by 
$\Omega_4=\delta^\bmu_\bnu \,\gamma^1_{\alpha\beta}$, 
$\Omega_5=\delta^\bmu_\bnu \,\gamma^2_{\alpha\beta}$ and 
$\Omega_s=iJ^{(s)\bmu}\!\,_\bnu \delta_{\alpha\beta}$ for  
$s=1,2,3$. The supersymmetry algebra again requires the manifold
to be hyperK\"ahler. As in the previous subsection, the $G^I$'s must
 be
triholomorphic Killing vector fields.

When quantized, the spinors $\psi^E = e_\bmu^E \psi^\bmu$ with vielbein 
$e_\bmu^E$, commute with all the bosonic dynamical variables, especially with  
$p$'s that are canonical momenta of the coordinate $z$'s. 
The remaining fundamental  commutation relations are  
\begin{eqnarray}  
&&[z^\bmu, p_\bnu ] = i\delta^\bmu_\bnu, \nonumber\\  
&&\{\psi^E_\alpha, \psi^F_\beta\} = \delta^{EF}\delta_{\alpha\beta}\,.  
\label{commutators}  
\end{eqnarray}  
Define  supercovariant momenta by  
\begin{eqnarray}  
&& \pi_\bmu \equiv p_\bmu -{i\over 2}\omega_{EF\,\bmu}  
\bar\psi^E \gamma^0 \psi^F,  
\label{cov}  
\end{eqnarray}  
where $\omega_{EF\,\bmu}$ is the spin connection. The N=4 SUSY generators 
in real spinors can be written as,  
\begin{eqnarray}  
&&Q_\alpha = \psi^\bmu_\alpha \pi_\bmu  
 - 
(\gamma^0\Omega^\bI \psi)^\bmu G^{\bI}_\bmu, 
\label{generator0} 
\\  
&&Q^{(s)}_\alpha = (J^{(s)}\psi)^{\bmu}_\alpha \pi_\bmu  
- 
(\gamma^0 J^{(s)}\Omega^\bI \psi)^\bmu G^{\bI}_\bmu\,.  
\label{generator}  
\end{eqnarray}  
These charges satisfy the $N=4$ complex superalgebra:  
\begin{eqnarray}  
&&\{Q_\alpha,Q_\beta  \}  =\{Q^{(s)}_\alpha,Q^{(s)}_\beta\}=2  
 \delta_{\alpha\beta} \; {\cal H}  
-2(\gamma^0\gamma^1)_{\alpha\beta} \; {\cal Z}_{4} 
-2(\gamma^0\gamma^2)_{\alpha\beta} \; {\cal Z}_{5}, \\  
&& \{Q_\alpha,\ Q^{(s)\,}_\beta  \} =  
2\gamma^0_{\alpha\beta} \; {\cal Z}_{s},\ \ \  
\{Q^{(1)}_\alpha,Q^{(2)}_\beta  \} = 
2\gamma^0_{\alpha\beta} \; {\cal Z}_{3},\\  
&& \{Q^{(2)}_\alpha,Q^{(3)}_\beta  \} = 
2\gamma^0_{\alpha\beta} \; {\cal Z}_{1}, 
\ \ \  \{Q^{(3)}_\alpha,Q^{(1)}_\beta  \} = 
2\gamma^0_{\alpha\beta} \; {\cal Z}_{2}\,,  
\end{eqnarray}  
where $\cal H$ is the Hamiltonian, and the ${\cal Z}_I$'s
are central charges,  
\begin{eqnarray}
{\cal Z}_{I}= G_{\bI}^{\bmu} \pi_\bmu - 
{i\over 2}  \nabla_\bmu G^{\bI}_\bnu \,\bar\psi^\bmu  
\gamma^0\psi^\bnu.   
\end{eqnarray} 
The sigma-model without the potential possesses an $SO(5)$ R-symmetry
which is explicitly broken by the $G^I$'s. The $G^I$'s transform as {\bf 5}
of $SO(5)_R$.
 
The complex form of the supercharges 
is often useful. To this end, we introduce  
$\varphi^\bmu\equiv{1\over \sqrt{2}} (\psi_1^\bmu-i\psi_2^\bmu)$ 
and define $Q\equiv {1\over \sqrt{2}}(Q_1-iQ_2)$. The supercharges 
in (\ref{generator0}) can be rewritten as 
\begin{eqnarray}  
&&Q = \varphi^\bmu \pi_\bmu  
 -\varphi^{*\bmu} (G^4_\bmu-iG^5_\bmu)-i\sum_{s=1}^3 
G^s_\bmu (J^{(s)}\varphi)^\bmu ,
\\  
&& 
Q^\dagger = \varphi^{*\bmu} \pi_\bmu  
 -\varphi^{\bmu} (G^4_\bmu+iG^5_\bmu)+ 
i\sum_{s=1}^3 G^s_\bmu (J^{(s)}\varphi^*)^\bmu 
\,. 
\label{generators}  
\end{eqnarray} 
The charges $Q^{(s)}$ and ${Q^{(s)}}^\dagger$ are analogously defined
from (\ref{generator}). 
The positive definite nature of the Hamiltonian can be seen easily 
in the anticommutator
\begin{equation}
\{Q,Q^\dagger\}=\{Q^{(s)},{Q^{(s)}}^\dagger\}=2{\cal H},
\end{equation}
while the central charges appear in other parts of the superalgebra. 
For instance, we have
\begin{eqnarray}
\{Q,Q\}=-{\cal Z}_4+i{\cal Z}_5, &&\nonumber \\
\{Q^\dagger,Q^\dagger\}=-{\cal Z}_4-i{\cal Z}_5. &&
\end{eqnarray}
Once we adopt this complex notation, it is natural to introduce an equivalent
geometrical notation. Defining the vacuum state $|0\rangle$ to be annihilated
by $\varphi^{*m}$'s, and using the 1-1 correspondence,
\begin{equation}\label{29}
(\varphi^{m_1}\varphi^{m_2}\cdots\varphi^{m_k})|0\rangle
\quad\leftrightarrow\quad dz^{m_1}\wedge dz^{m_2}\wedge 
\cdots \wedge dz^{m_k},
\end{equation}
we can reinterpret $\varphi^{m}$ as the exterior product with $dz^m$,
and $\varphi^{*}_{m}=g_{mn}\varphi^{*n}$ as the contraction with 
$\partial/\partial z^m$. The supercharge operators can be rewritten as,
\begin{eqnarray}
Q&=&
-id- \iota_{G^4-iG^5} +i\iota_{J^{(1)}(G^1)}^\dagger
+i\iota_{J^{(2)}(G^2)}^\dagger + i\iota_{J^{(3)}(G^3)}^\dagger, \nonumber \\
Q^\dagger &=& 
id^\dagger -\iota_{G^4+iG^5}^\dagger - i\iota_{J^{(1)}(G^1)}
-i\iota_{J^{(2)}(G^2)} - i\iota_{J^{(3)}(G^3)},
\end{eqnarray}
where $\iota_K$ is the contraction with the vector field $K$, and its conjugate
$\iota_K^\dagger$ is the exterior product by the 1-form obtained from $K$ by
lowering its indices.

The SUSY quantum mechanics admit a canonical $Z_2$ grading, which in the 
geometrical notation of (\ref{29}) is defined on $k$-forms by 
\begin{equation}
\tau_4\equiv(-1)^k.
\end{equation}
or equivalently by
\begin{equation}
\tau_4\equiv \prod 2\psi^E_1\psi^E_2 =  
\prod (\varphi^{*E}\varphi^{E}-\varphi^{E}\varphi^{*E}) .
\end{equation}
The involution $\tau_4$ anticommutes with all supercharges and 
determines the usual Witten index, ${\cal I}_4$. 

In some special limits, however, 
there could be an additional $Z_2$ grading. Suppose that we have only one 
nonzero $G^I$, say $G^5$. The operators
\begin{equation}
\tau_\pm \equiv\prod (\sqrt{i}\,\varphi^E\pm \sqrt{-i}\,\varphi^{*E}).
\end{equation}
anticommutes with the Dirac operators defined as,
\begin{equation}
D_\pm\equiv iQ \pm Q^\dagger= (i\varphi^m \pm \varphi^{*m})(\pi_m \mp G^5_m)
= d- \iota_{G^5} \pm i(d^\dagger- \iota_{G^5}^\dagger),
\end{equation}
the square of which is
\begin{equation}
D_\pm^2=\pm 2i({\cal H}\mp {\cal Z}_5).
\end{equation}
So the $Z_2$ gradings define an analog of the signature index for each choice
of sign and for each charge-eigensector. A given state with nonzero 
${\cal Z}_5$ can be annihilated by one of $D_\pm$ at most, and in fact 
must break at least half of the supercharges. The corresponding indices
will be denoted by ${\cal I}_s^\pm$.

In Section 5, we will compute both ${\cal I}_4$ and ${\cal I}_s^\pm$
in such a special limit with only one of five $G^I$'s present, which 
we can take to be $G^5$ without loss of generality. For ${\cal I}_4$,
we may take any one of $D_\pm$ as the Dirac operator, since $\tau_4$
anticommutes with both. A standard
index theorem will then allow us to deduce ${\cal I}_4$ in more
general setting.

\section{Moduli Spaces}

Moduli space dynamics of monopoles decompose into the interacting relative 
part and the non-interacting ``center of mass'' part. The latter corresponds 
to a 4-dimensional flat metric of the form,
\begin{equation}
g_{cm}= A\, d\vec X^2 + B\,d\xi_T^2,
\end{equation}
where $\vec X$ is a three-vector. Since we are interested in establishing 
existence of bound states, this part of the dynamics will be ignored for 
the most part. 

The free center-of-mass sector generates two kinds of quantum numbers, 
nevertheless. One is the overall, conserved $U(1)$ charge, and the other
is a supermultiplet structure generated by the fermionic partners
of $\vec X$ and $\xi_T$. The resulting degeneracies, 4 and 16 for $N=4$
real and complex supersymmetric quantum mechanics respectively, correspond
to the smallest possible BPS multiplet of the underlying SUSY Yang-Mills 
field theories with 8 and 16 supercharges, respectively.

\subsection{Distinct Monopoles}

A simple case of this dynamics involves a collection 
of distinct monopoles in $SU(n)$ gauge theories. The interacting part of the
moduli space metric is a simple generalization of four-dimensional 
Taub-NUT metric \cite{many}. 
Without loss of generality, consider a collection of 
$k+1$ distinct monopoles, whose magnetic charges are given by an 
irreducible (sub)set of simple roots, $\beta_1,\dots,\beta_{k+1}$.
The simple roots satisfy relations $\beta_a^2=1$, 
$\beta_a\cdot\beta_{a+1}=-1/2$, and $ \beta_a\cdot\beta_{a+b}=0$ for $b>1$.
The relative part of the corresponding metric is
\begin{equation}
g=C_{ab}\;d\vec r_a\cdot d\vec r_b +\frac{4\pi^2}{e^4}(C^{-1})_{ab}
(d\psi_a+\cos\theta_a d\phi_a)
(d\psi_b+\cos\theta_b d\phi_b),
\end{equation}
where the matrix $C$ for the relative moduli space is\footnote{The coupling
constant $e$ will be assumed to be positive without loss of generality.}
\begin{equation}
C_{ab}=\mu_{ab}+\frac{2\pi}{e^2}\delta_{ab}\,\frac{1}{r_a}.
\end{equation}
The 3-vector $\vec r_a$ is the relative position between the $a^{th}$ and 
$(a+1)^{th}$ monopoles, 
\begin{equation}
\vec{r}_a = {\vec x}_{a+1}-{\vec x}_{a},
\end{equation}
while the angles $\psi_a$ of period $4\pi$ are related to the $U(1)$
phases of each monopole, $\xi_a$'s (of period $2\pi$), by
\begin{eqnarray}
2\frac{\partial}{\partial \psi_a}&=&\frac{\partial}{\partial \xi_{a+1}}
- \frac{\partial}{\partial \xi_a}\nn
\left(\sum_{a=1}^{k+1}m_a\right)\frac{\partial}{\partial\xi_T}&=&
\sum_{a=1}^{k+1} m_a\frac{\partial}{\partial\xi_a}.
\end{eqnarray}
where $\xi_T$ is a coordinate that appears in free center-of-mass part of 
the dynamics and $m_a$ is the mass of the $a^{th}$ monopole. 

For a generic reduced mass matrix $\mu$, the  triholomorphic Killing 
vector fields of this geometry are exhausted by
\begin{equation}
K_a=\frac{\partial}{\partial \psi_a},
\end{equation}
so the vector fields $G$ and $G^I$ are linear combinations of $K_a$'s
with constant coefficients;
\begin{eqnarray}
G&=&e\sum_c a_cK_c,\nonumber \\
G^I&=& e\sum_c a^I_c K_c.
\end{eqnarray}
The electric charges are measured by the charge operators,
\begin{equation}
-i{\cal L}_{K_a},
\end{equation}
whose (half-)integer eigenvalues will be denoted by $q_a$. In terms of the
simple roots $\beta_a$, the electric charge of a dyonic state with charge
$q_a$'s is
\begin{equation}
\begin{array}{l}
e(+q_1+q_2+q_3+\cdots+q_k+n/2)\beta_1 +\\
e(-q_1+q_2+q_3+\cdots+q_k+n/2)\beta_2 +\\
e(-q_1-q_2+q_3+\cdots+q_k+n/2)\beta_3 +\\
e(-q_1-q_2-q_3+\cdots+q_k+n/2)\beta_4 +\\
\vdots\\
e(-q_1-q_2-q_3-\cdots-q_k+n/2)\beta_{k+1}.\\ 
\end{array}
\end{equation}
where the integer $n$ comes from quantization of an overall $U(1)$ angle
and should be even or odd when $2\sum_a q_a$ is even or odd, respectively.

\subsection{Unit Noncommutative Instanton}

A simple deformation of the above moduli space appeared in another
context recently, where one considers low energy dynamics of an
instanton soliton in the 5-dimensional noncommutative $U(k+1)$ Yang-Mills 
theory \cite{nek}. This happens because an instanton in $S^1\times R^3$ 
can be regarded as a collection of $k+1$ distinct monopoles of the
underlying Yang-Mills theory \cite{caloron}.
When we compactify the theory on a circle of radius $R$, the
nontrivial part of the moduli space of a single instanton soliton is 
given by the metric \cite{nci}
\begin{equation}
g=\frac{4\pi^2R}{\tilde e^2}
\left(\tilde C_{ab}\;d\vec r_a\cdot d\vec r_b +(\tilde C^{-1})_{ab}
(d\psi_a+\cos\theta_a d\phi_a)
(d\psi_b+\cos\theta_b d\phi_b)\right),
\end{equation}
where $\tilde e$ is the 5-dimensional Yang-Mills coupling.
The matrix $\tilde C$ for the relative moduli space is
\begin{equation}
\tilde C_{ab}=\nu_{ab}+\delta_{ab}\,\frac{1}{r_a} 
+\frac{1}{|\sum \vec r_a-2\pi\vec\zeta/R\,|} ,
\end{equation}
where $\vec \zeta$ encodes the noncommutativity. The matrix $\nu$ is 
determined by the Wilson line along $S^1$ that breaks the
 gauge symmetry to $U(1)^n$. When the supersymmetry
of the underlying field theory is maximal with 16 supercharges, the 
low energy dynamics of the instanton is given by our SUSY quantum
mechanics with 4 complex supercharges. When the field theory
comes with 8 supercharges, the instanton dynamics is described
by the SUSY quantum mechanics with 4 real supercharges.

\section{Bound States of Two Distinct Monopoles}

For a pair of two distinct and interacting monopoles, the dynamics have 
been solved for supersymmetric ground states in each charge eigensector. 
The geometry is that of a Taub-NUT manifold which comes with a single
triholomorphic Killing vector field $K_1$. Accordingly, there is only 
one conserved $U(1)$ charge, $q_1$, which labels superselection sectors. 

In the pure $N=2$ Yang-Mills case, define $\tilde a_1 \equiv 4\pi^2 a_1/e^3\mu$
where $\mu$ is the reduced mass and $a_1$ is defined by 
$G=e\,a_1K_1$. The normalizable wavefunctions had been constructed
by Pope in another context \cite{pope}, and the number of
dyonic bound states of charge $q_1$ was found to be \cite{n2}
\begin{equation}
2|q_1|,
\end{equation}
if $0<q_1<\tilde a$ or $\tilde a_1<q_1<0$, and 
\begin{equation}
0,
\end{equation}
otherwise. For each $q_1$,
the solutions belong to the same chirality spinors, and thus contribute
to the Witten index equally. Thus the Witten index ${\cal I}_2$ in each
charge eigensectors are
\begin{equation}
{\cal I}_2= \left\{\begin{array}{cl} 2|q_1| & 
\qquad 0<|q_1|<|\tilde a_1|\quad\hbox{and}\quad 0< q\tilde a_1 \\
&\\
0 &\qquad {\rm otherwise}\end{array} \right\}.
\end{equation}
For a pair of distinct monopoles in $N=4$ Yang-Mills \cite{1/4}, 
the five $G^I$'s must be proportional to the single triholomorphic 
vector field $K_1=\partial/
\partial\psi_1$. We may rotate them into a single triholomorphic vector 
field, say $G^{I=5}=e\,a_1K_1$, upon which we can define $\tilde a_1$ 
similarly as 
above, $\tilde a_1 \equiv 4\pi^2 a_1/e^3\mu$. The degeneracy is found to be
\begin{equation}
1,
\end{equation}
for purely magnetic state ($q_1=0$), while for dyons
\begin{equation}
8|q_1|,
\end{equation}
when $0< |q_1| < |\tilde a_1|$, 
and zero otherwise. All solutions are self-dual
differential forms, when we take the convention that the curvature tensor
of the moduli space is self-dual. 

When the central charge ${\cal Z}_5=ea_1q_1$ of the state is positive
(negative), the bound state is annihilated by $D_+$ ($D_-$) only, while for 
${\cal Z}_5=0$, the state is annihilated by both. For given $q_1$, we find
\begin{equation}
{\cal I}_s^+=
\left\{\begin{array}{cl} 
1& \qquad q_1=0 \\
8|q_1| & \qquad 0< |q_1| < |\tilde a_1| \quad{\rm and} \quad q_1a_1>0\\
0 &\qquad {\rm otherwise}\end{array} \right\} ,
\end{equation}
and 
\begin{equation}
{\cal I}_s^-=
\left\{\begin{array}{cl} 
1& \qquad q_1=0 \\
8|q_1| & \qquad 0< |q_1| < |\tilde a_1| \quad{\rm and} \quad q_1a_1 <0\\
0 &\qquad {\rm otherwise}\end{array} \right\} .
\end{equation}
The Witten index ${\cal I}_4$ counts the number of even 
forms minus the number of the odd forms. Of solutions with $q_1\neq 0$, 
half are even and the other half are odd, so we find that
\begin{equation}
{\cal I}_4= \left\{\begin{array}{cl} 
1 & \qquad q_1=0 \\
&\\
0 &\qquad q_1\neq 0\end{array} \right\}
\end{equation}
regardless of $a_1$.

\section{Index Computation}

We would like to put a lower bound on the number of bound states in the above
SUSY
dynamics by computing indices. The index problems can be quite 
involved, given that the quantum mechanics involve many degrees of freedom 
with complicated interaction terms. However, the problem can be simplified 
by utilizing the invariance of the index under
certain deformations. In this section we will use the invariance 
of the index under Fredholm deformation to simplify our index computations. 
Before proceeding with the computation, however, we need to restrict to 
the regime where a massgap exists.

\subsection{Massgap}

When restricted to specific charge 
eigensectors, the operators above may exhibit two drastically different
behavior. For small charges, the sector has a massgap; the continuum
part of the spectrum is bounded below by a positive gap. For large
charges, the massgap disappears. This is the reason why there is an
upper bound on the electric charge $q_1$ of bound states of two monopoles.
In the two-body problems, the condition for the massgap to exist in a 
sector with electric charge $q_1$ is
\begin{equation}
|q_1| < \frac{4\pi^2}{e^3}\frac{|a_1|}{\mu},
\end{equation}
where the bosonic potential is generated by a single triholomorphic
vector field $G=e\, a_1 K_1$. 
When we consider many distinct monopoles, the condition for the massgap to
exist is equally simple:
\begin{equation}
|q_c| <  | \tilde a_c |,
\end{equation}
where
\begin{equation}
\tilde a_c \equiv
\frac{4\pi^2}{e^3}  \sum_{b=1}^k (\mu^{-1})_{cb}\,a_b ,
\end{equation}
with $G=ea_cK_c$. In the quantum mechanics with four complex 
supersymmetries, the same holds true provided that only one $G^I$, say, 
$G^5=ea_cK_c$ is 
turned on. We will compute the indices, ${\cal I}_2$, ${\cal I}_4$, 
${\cal I}_s^\pm$ assuming that all of these conditions hold.\footnote{If  
five $G^I=e\sum_c a^I_c K_c$'s are involved, the massgap 
condition generalizes to
\begin{equation}
(q_c)^2 <  \sum_{I=1}^5 (\tilde a^I_c)^2 ,
\end{equation}
where $\tilde a^I_c$ are defined similarly as above for each $G^I$. However,
the Indices ${\cal I}_s^\pm$ are not well-defined unless all $G^I$'s are
proportional to each other. We will discuss such generic cases 
in Section 6.}

\subsection{Index Generalities}

First we recall basic definitions. 
\begin{define}
A bounded linear operator $L:E_1\rightarrow E_2$ between two Hilbert spaces
is {\em Fredholm} if there exists a bounded operator  
 $P :E_2\rightarrow E_1$ 
such that 
$PL-I_1$, and $LP-I_2$ are compact operators. Here $I_j$ denotes the identity 
map on $E_j$.
\end{define}

The operator $P$ in the above definition is called a parametrix. 
We will be interested in the case where $L$ is a Dirac operator. 
In this case, although Dirac operators are unbounded on $L_2$, we
 may trivially 
make $L$ bounded by taking $E_1$ to be the closure of the domain of $L$ 
with respect to the norm (graph norm) 
$$\|f\|_{graph}^2 \equiv \|f\|^2 + \|Lf\|^2,$$
where unsubscripted norms denote $L_2$ norms. 

If $L$ is a Dirac operator on a compact manifold, then it is well known 
to be Fredholm. In the compact case, one takes, for example, 
$P$ to be the Green's operator, ${\cal G}$ defined to be the unique 
operator satisfying:\newline 
(i) ${\cal G}$ annihilates the kernel of $L^\dagger$.\newline
(ii) The range of ${\cal G}$ is orthogonal to the kernel of $L$.\newline
(iii) $L{\cal G}f = f$ for $f\in$ the image of $L$.

Then  ${\cal G}$ is bounded by $1+\lambda_1^{-1/2}$, where $\lambda_1$  is 
the first nonzero eigenvalue of $L^\dagger L.$ Also, 
${\cal G}L = I_1-\Pi_1,$ and $L{\cal G}=I_2-\Pi_2$, 
where $\Pi_1$ and $\Pi_2$ denote the orthogonal projections onto the kernels 
of $L$ and $L^\dagger$ respectively (and are finite rank and thus 
compact operators). Hence 
$P={\cal G}$ satisfies all the conditions of the definition.

In the case of a Dirac operator on a noncompact 
manifold the preceding construction of a Greens operator may 
fail to yield Fredholmness for 
several reasons. The kernel of $L$ or $L^\dagger$ may fail to be finite
 dimensional, 
making one of the projections not a compact operator. Also, if there is no
gap in the spectrum, ${\cal G}$ 
will fail to be bounded. These deficits are all 
avoided, however, under the assumption that 
the essential spectrum  
 of $L^\dagger L$  
is bounded away from zero. (We recall that the essential spectrum 
includes the continuous spectrum and any eigenvalue of infinite multiplicity.)
 Then the kernels are finite dimensional 
and ${\cal G}$ is again bounded by  $1+\lambda_1^{-1/2}$, where
 $\lambda_1$ is 
the smallest nonzero element of the spectrum of $L^\dagger L$. 
  It is well known that the essential spectrum 
is bounded away from zero whenever $L^\dagger L$ has the form 
$\Delta + V$, where for two positive constants $c$ and $R$, $V$ satisfies  
$V(x) > c$ for $x$ outside a fixed compact set. All the operators 
we consider in this paper have this form.  

A basic result in index theory (eg \cite{palais}{ p.122}), is the following. 

\begin{prop}
Let $L_t$, $t\in [0,1]$ be a continuous family of Fredholm operators.
 Then $index(L_0) = index(L_1)$.  
\end{prop}

Thus one can sometimes deform an index problem to a more 
tractable index computation. To avoid potential confusion, 
we recall the notion of continuity assumed in the above proposition.
 $L_t$ is a continuous family of operators if for each $s$ and 
for every $\epsilon > 0$, there exists $\delta >0$ so that 
$\|L_t F - L_sF\|/\|F\| < \epsilon$ for all nonzero $F$ 
if $|t-s|<\delta$. In particular, we note that we require the 
$\delta$ to be independent of $F$. Hence, for example,
 the super harmonic oscillator in one variable 
$\psi_1 \frac{d}{dx} + \psi_2 x$, cannot be continuously deformed 
to $\psi_1 \frac{d}{dx} + \psi_2$ by scaling away the interaction term. 
If one uses the graph norm for $\psi_1 \frac{d}{dx} + \psi_2$, then 
 the oscillator is unbounded and hence clearly cannot be deformed 
to a bounded operator. If one instead 
uses the graph norm for the oscillator, 
 it is easy to see that all frequencies give equivalent norms and  
by construction, multiplication by $x$ (as a map to $E_2$ 
is continuous in each of these norms. 
Hence the deformation 
$$L_t = \psi_1 \frac{d}{dx} + \psi_2 ((1-t)x+t)$$
is continuous. The limit operator, however, is not Fredholm as a 
map from $E_1\rightarrow E_2$ even though it is easy to show that it is 
Fredholm if the oscillator graph norm is replaced by the
 $\psi_1 \frac{d}{dx} + \psi_2$ graph norm. 

In analyzing continuous families of operators $L_t$ it is often useful to
utilize also families of parametrices $P_t$. If, however, 
 we choose $P_t$ to be
the Greens operator ${\cal G}_t$ of $L_t$ then we will
 be plagued by the possibility that if eigenvalues converge to zero,
${\cal G}_t$ will become unbounded. Hence, 
for no other reason than to avoid such 
problems of bounding $P_t$, it is useful to define a modified 
Greens operator 
$${\cal G}_{L_t,\epsilon} = {\cal G}_t(I-\Pi_{L_t,\epsilon}),$$
where $\Pi_{L_t,e}$ denotes the projection onto the $\lambda\leq \epsilon$ 
eigenspaces of $L^\dagger_tL_t$. This operator is bounded by 
$1+\epsilon^{-1/2}$ 
and is a parametrix as long as $\epsilon$ lies below the essential spectrum.

\subsection{Deforming the Index}

For several of the index problems we will be considering, it seems likely 
that one can simply deform the given operator into a standard superharmonic 
oscillator and then immediately deduce the index. There are some minor 
issues fitting such a deformation into a continuous family. We 
will not treat those here because for one of our index computations
 - that of the noncommutative instanton - there is no single model operator 
to which to deform. Instead we will use the deformation invariance 
of the index to localize all the problems to an elementary
 computation around the zeros 
of our triholomorphic vector field $G$. 

The case of interest to us then is $D$ a Dirac operator 
anticommuting with an involution $\tau$, 
$L$ the restriction of $D$ to the $+1$ eigenspace of $\tau$,
and $E_1$ and $E_2$ the spaces of sections of the associated bundles
with finite graph and $L_2$ norms respectively.   
In this context, Fredholmness follows from the conditions in the preceding 
sections guaranteeing a mass gap (i.e., bounding the essential 
spectrum of $D^\dagger D$ away from $0$.) 

The deformations we will consider involve replacing $G$ by $TG$ for some 
$T$ large. 
This fits into the above framework without modification since $T\geq 1$ 
ensures preservation of the mass gap. Moreover, scaling $G$ is clearly 
continuous because the norm of $G$ is a bounded in the given metric. 
Recall that $G$ enters the Dirac operators in the form of operators,
\begin{equation}
\lambda^m G_m 
\end{equation}
or
\begin{equation}
(\sqrt{i}\,\phi^m \pm \sqrt{-i}\,\phi^{*m})G_m 
\end{equation}
which are Clifford multiplications by $G$. Denote these operators by $\hat G$. 
We see that the sup norm
of the difference between two Dirac operators (associated to $TG$ and $SG$) 
is bounded by 
$\|(T\hat G-S\hat G)f\| \leq |(T-S)|\times |G|_{sup}\times\|f\|,$
which clearly gives the desired inequalities for the continuity of the
deformation. We note that even had the metric allowed for unbounded $|G|$, 
we still would have  $\hat G$ bounded as an operator from 
$E_1$ (equipped with the graph norm) to $E_2$, as in the
oscillator example of the previous section. 
 
In addition, we will modify the metric on certain compact
 subsets. This modification may change the actual domain 
and range of our operator. For example $\tau$, and hence its eigenspaces 
may vary with the metric. Nonetheless, we may choose quasiisometries
 between them. Thus if we have Fredholm operators 
$D_T:E_1(T)\rightarrow E_2(T)$ and 
quasiisometries $h_i(T):E_i\rightarrow E_i(T)$ then 
the index of $h_2(T)^{-1}D_Th_1(T)$ is $T$ independent by the proposition 
and is equal to the index of $D_T$ since the index is unchanged 
under composition with bounded operators with bounded inverse.  
  We note, although we will not need it here, that the condition that 
$h_i$ be quasiisometries may be relaxed to the condition that the 
eigenvalues of $h_i$ and $h_i^{-1}$ grow at most polynomially 
(subexponentially even) in distance 
from some choice of origin. This is an easy consequence of the fact that the 
Fredholm estimate implies exponential decay of the elements in the 
$L_2$ kernel of $D_T$. (See \cite{agmon}). 
As we will use these decay properties, let us recall them in a crude form 
now.         
  
Suppose  we have $N$ points $y_i, i=1,\cdots, N$ and 
a Hamiltonian of the form 
$H = \Delta + 4T^2V,$ with $V(x) \geq 1 $ if $|x-y_i| > 1$, $i=1,\cdots,N$.  
Suppose also that $Hf = \lambda_0^2f$ for some small constant $\lambda_0$, 
and $f\in L_2$, say with $L_2$ norm $1.$
Let 
$$|x|_m :=min_{1\leq i\leq N}|x-y_i|.$$
 Then  $e^{(T^2-\lambda_0^2)^{1/2}|x|_m}f\in L_2,$
and the $L_2$ norm of $|e^{(T^2-\lambda_0^2)^{1/2}|x|_m}f|$ 
restricted to the exterior of the balls of radius $R>1$ about the $y_i$ is 
finite and bounded by $4e^{(T^2-\lambda_0^2)^{1/2}R}$. 
(This is not sharp. See \cite{agmon} for sharper statements.) In particular, 
we observe that the $L_2$ norm of $f$ restricted to the complement of the balls 
of radius $2R$ about the $y_i$ satisfies
\begin{equation}\label{est1} 
\|f_{|{B_{2R}^c}}\|^2\leq 4e^{-2(T^2-\lambda_0^2)^{1/2}R}.
\end{equation}
Hence, $f$ is concentrated near the zeroes of $V$.

Let $D_T$ denote our Dirac operator with $G$ replaced by $2TG$ and 
the metric modified to be Euclidean in a ball of radius $10R$ some $R>>1$
about each zero of $|G|^2$.  This metric modification allows us to 
compare $D_T$ to a model Dirac operator which agrees with 
$D_T$ near the zeros of $G$ and has known index. 
Assume, as we may by replacing  $G$ initially by a suitable 
multiple, that  $D_T^*D_T$ has the form 
$\Delta + 4T^2V$, with $V(x)>1$ in the 
complement of the balls of radius 1 about each zero of $G$.
As in the previous section, for $\epsilon$ below the continuous spectrum of
$D_T^2$, $\Pi_{D_T,\epsilon}$ denotes the projection onto the 
$\lambda \leq \epsilon$ eigenspaces of $D_T^2$. Then 
\begin{equation}\label{int}
{\rm Index} D_T^+ = \int dx\; {\rm tr}\;\tau \Pi_{T,\epsilon}(x,x).
\end{equation} 
Using (\ref{est1}) we see that 
for $d_\epsilon := {\rm rank }\,\Pi_{D_T,\epsilon}$,
\begin{equation}\label{est3}
\int dx\;{\rm tr}\;\tau \Pi_{D_T,\epsilon}(x,x) = \int_{|x|_m < 2c}dx\;
{\rm tr}\;\tau \Pi_{D_T,\epsilon}(x,x)
+ O(d_\epsilon e^{-2c(T^2-\epsilon)^{1/2}}),
\end{equation} 
for some choice of $c$. 
Hence it suffices to bound $d_\epsilon$ independently of $T$ large and 
to compute the integral of ${\rm tr}\;\tau \Pi_{T,\epsilon}(x,x)$ 
over $|x|_m<2c$ in the large  $T$ limit. 

First we estimate $d_\epsilon$. Let $D_T^2f = \lambda^2f,$ for 
$\lambda\leq \epsilon$, and 
$ \|f\|_{L_2} = 1.$
Let $Q_T$ denote the Green's operator for the super harmonic oscillator (SHO) 
which agrees with $D_T^2$ in a neighborhood of radius $4R$ about 
the zeros of $G$. 
Let $\rho_R$ denote a cutoff function supported on a ball of radius $2R$ 
where $D_T^2$ is the SHO and identically one 
on a ball of radius $R$. Then 
$(\rho_R f - Q_TD_T^2 \rho_R f)$ is in the kernel of the SHO. Denote its
norm by $a$ and introduce a unit vector $v$ in the kernel of SHO such that
$$(\rho_R f - Q_TD_T^2 \rho_R f) = av.$$
Observe that $Q_T$ has sup norm $\leq  T^{-1}.$ 
Now consider the equality 
$$D_T^2 \rho_R f = \lambda^2\rho_R f + [\Delta,\rho_R]f.$$
By our assumptions, the right hand side is $O(\lambda^2) + O(e^{-TR})$ 
(not sharp). 
Hence 
$\|\rho_R f - av\| = O(\lambda^2) $ for $\lambda > O(e^{-TR/2}).$
Setting, for example, $\epsilon = 1/T,$ we have then 
\begin{equation}
\|f - av\| = O(1/T^2). 
\end{equation}
Moreover, such an inequality is true for any vector in the image of 
$\Pi_{D_T,1/T}$. 
We  conclude then that rank $\Pi_{D_T,1/T}$ is no larger than the dimension 
of the kernel of the SHO (times the number of zeros of $G$). 
This bounds $d_{1/T}$ and completes our demonstration that it suffices to 
compute the trace over a bounded region. 
 
In the following, for simplicity of notation we will consider the case of 
a single zero for $G$, but the general case follows similarly with 
only notational complications. 

Let $S_T^+$ and $\tau_E$ denote the Euclidean Dirac operator and involution 
which agree with $D_T^+$ and $\tau$ near the zeros of $G$. Let $F_T$ denote 
the Greens operators for $S_T^+$. To define a parametrix for $D_T$, introduce
${\cal G}_T$, the Greens operator for $D_T^+$, and let $P_T$ be
the modified Greens operator
$$P_T:={\cal G}_T(I-\Pi_{D_T,1/T}).$$

Define   
$$I_1:={\rm Index}(D_T^+)-{\rm Index}(S_T^+) = Tr([D_T^+P_T -S_T^+F_T] - 
[P_TD_T^+ - F_TS_T^+]).$$
Then we wish to show that the integer $I_1 = 0$. 
By (\ref{est3}), the above traces can be approximated for $R,T$ large as 
$$I_1 = {\rm Tr}\rho_R ([D_T^+P_T -S_T^+F_T] - [P_TD_T^+ - F_TS_T^+]) + 
O(d_{1/T} e^{-RT/2}).$$
On the support of $\rho_R,$ $D_T^+=S_T^+,$ hence we have
$$I_1 = {\rm Tr}\rho_R [D_T^+,P_T -F_T] + O(d_{1/T} e^{-RT/2}).$$ 
Using the cyclic property of the trace, we rewrite the first term on the 
right hand side of the above formula as 
$${\rm Tr}\rho_R [D_T^+,P_T -F_T] = -Tr[D^+,\rho_R](P_T -F_T) + 
\int dx\; \nabla_iV(x)^i.$$
where $V(x)$ is the vector with 
$$V_i(x):= {\rm Tr}\gamma_i(\rho_R(x)(P_T -F_T)(x,x)).$$
The integral vanishes by Stoke's theorem, leaving 
$$I_1 = -\int {\rm tr}[D^+,\rho_R](x)(P_T -F_T)(x,x)dx.$$
We estimate the last term by converting it back into an expression involving 
the exponentially decaying projection operators. 
Write
\begin{eqnarray}
&&[D_T^+,\rho_R](P_T -F_T) \nn
= && 
[D_T^+,\rho_R](F_TS_T^+ + \Pi_{S_T^+,1/T})(P_T -F_T) \nn
=&&[D_T^+,\rho_R]F_T(S_T^+P_T -S_T^+F_T) + 
[D_T^+,\rho_R](\Pi_{S_T^+,1/T})(P_T -F_T).
\end{eqnarray}
The last term is  $O(Te^{-RT})$ because 
$$[D_T^+,\rho_R](\Pi_{S_T^+,1/T})=O(e^{-RT})$$ 
by (\ref{est1}) and because $(P_T - F_T)$ has sup norm $< T$ by construction. 
We separate the first term into two additional terms 
\begin{eqnarray}
&&[D_T^+,\rho_R]F_T(S_T^+P_T -S_T^+F_T)  \nn
=&&[D_T^+,\rho_R]F_T(D_T^+P_T-S_T^+F_T) + 
[D^+,\rho_R]F_T(S_T^+-D_T^+)P_T.
\end{eqnarray}
The first term is again exponentially decreasing because 
$(D_T^+P_T-S_T^+F_T)$ is a difference of exponentially decaying 
projection operators and 
$F_T$ is uniformly bounded.  We can compute $F_T$ explicitly and it 
is $O(e^{-T\delta})$, where $\delta$ is the distance between 
the support of $[D_T^+,\rho_R]$ and the support of $(S_T^+-D_T^+)$.
 Hence all the terms are exponentially decreasing, and we deduce that 
$I_1$ is exponentially decreasing. 
On the other hand, it is the difference between two integers 
and must therefore 
vanish.   

We summarize our results. 
$${\rm Index}(D_T^+) = {\rm Index}(S_T^+).$$
When there is more than one zero of $G$, a minor variation of 
the same argument yields 
$${\rm Index}(D_T^+) = \sum_i {\rm Index}(S_T(i)^+),$$
where $S_T(i)$ is the local model for $D_T$ at the $i^{th}$ zero of $G$. 
In order to extend our argument to this case, 
we must replace expressions of the form $F_TP_T$ and $\Pi_{S_T^+,1/T}P_T$
 in the previous expression 
by $F_T\rho_{nT}P_T$ and $\Pi_{S_T^+,1/T}\rho_{nT}P_T$ for some large $n$, 
because $F_T$ and $\Pi_{S_T^+,1/T}$ need not extend naturally to the 
full moduli space in the many zero case. This will introduce new error 
terms  of the form $\nabla\rho_{nT}P_T\rho_T to be estimated.$ 
Our decay estimates can once again be used to show these terms are also 
exponentially decaying.

\subsection{Computing the Deformed Index} 
We now use the deformation arguments of the preceding section 
to complete the index computations.
We consider first the case of the quantum mechanics with 4 complex 
supersymmetries on a moduli space of dimension 4k and compute ${\cal I}_s^+.$
The other cases are very similar and follow with minor modifications.
In the ${\cal I}_s^+$ case, we have reduced the problem to computing the 
index of the operator 
$S_{1/e}:= d-\iota_G + i(d^\dagger - 
\iota_G^\dagger)$ acting on selfdual forms 
($\tau_+  = 1$) 
on $C^{2k}$. Separating variables, we see that the index of $S_1$ 
is the product of the indices of the $D_c$, $c=1,\cdots,k,$ 
where where 
$D_c:=  d-\iota_{a_cK_c} + i(d^\dagger - 
\iota_{a_cK_c}^\dagger)$ (no sum over $c$) 
acting on selfdual forms on $C^2$. 
Using the deformation invariance of the index again, we may assume 
$a_c = 2$.  

This latter index is easy to calculate exactly as follows. 
We compute 
\begin{eqnarray}
&&D_cD_c^\dagger + D_c^\dagger D_c \nn
&&=\Delta + |2K_c|^2 - \{d,\iota_{2K_c}^\dagger\} + i\{\iota_{2K_c},d\} 
- i\{d^\dagger,\iota_{2K_c}^\dagger\} - \{d^\dagger, \iota_{2K_c}\}.
\end{eqnarray}
Let $z_1$ and $z_2$ denote complex coordinates on $C^2$.  
Then $2K_c = i\sum_{j=1}^2(z_c\frac{\partial}{\partial z_c}
 - \bar{z}_c\frac{\partial}{\partial \bar{z}_c}).$
Hence, $|2K_c|^2 = |z_1|^2+|z_2|^2.$ 
In the coordinate frame, on $(p,q)$ forms we have 
$$i\{\iota_{2K_c},d\}=-i\{d^\dagger,\iota_{2K_c}^\dagger\} 
= i{\cal L}_{2K_c} = -(p-q)-2K_c/i,$$
and 
\begin{eqnarray}
&&\{d,\iota_{2K_c}^\dagger\} = idz_1 d\bar{z}_1 + idz_2 d\bar{z}_2, \nn
&&\{d^\dagger,\iota_{2K_c}\} = idz_1^\dagger d\bar{z}_1^\dagger 
+ idz_2^\dagger d\bar{z}_2^\dagger.
\end{eqnarray}
Hence, we have 
\begin{eqnarray}
&&D_cD_c^\dagger + D_c^\dagger D_c \nn
=&& 
\Delta + |z|^2 -2(p-q)-4K_c/i - \{d,\iota_{2K_c}^\dagger\}
 - \{d^\dagger, \iota_{2K_c}\}.
\end{eqnarray}
The  functions
\begin{equation}
f(a,b,c,d):=(\partial_{z_1}-\bar{z_1})^a(\partial_{\bar z_1} - z_1)^b 
(\partial_{z_2}-\bar{z_2})^c(\partial_{\bar z_2} - z_2)^de^{-|z|^2/2},
\end{equation}
with $ a,b,c,d\geq 0,$ span the eigenspace of $\Delta + |z|^2 $ with 
the eigenvalue $2(a+b+c+d)+4.$ 
We compute commutators to obtain 
 $$2K_c f(a,b,c,d) = i(-a+b-c+d)f(a,b,c,d).$$
Therefore, on the algebraic span of $f(a,b,c,d)$ we have 
\begin{eqnarray}
D_cD_c^\dagger + D_c^\dagger D_c 
= 4(a+c) + 4 - 2(p-q) - \{d,\iota_{2K_c}^\dagger\} 
- \{d^\dagger,\iota_{2K_c}\}.
\end{eqnarray}
This vanishes if and only if $a=c=0$ and the differential form coefficient 
of $f(a,b,c,d)$ takes one of the following forms:
\begin{equation}
dz_1 \wedge dz_2,
\end{equation}
or 
$$ 1 + (idz_1\wedge d\bar{z}_1 + idz_2 \wedge d\bar{z}_2)/2 
- dz_1 \wedge d\bar{z}_1 \wedge dz_2 \wedge d\bar{z}_2,$$
or
$$(dz_1 + idz_2 \wedge d\bar{z}_2\wedge dz_1/2), $$
or
$$(dz_2 + idz_1 \wedge d\bar{z}_1\wedge dz_2/2).$$

We have, therefore, an infinite dimensional kernel to our operator
before taking into account the constraint on charge. We now recall that we 
wish to restrict to the space 
$$2q_c = {\cal L}_{2K_c}/i = (p-q)+2K_c/i.$$ 
With the above normalization of $K_c$, $q_c$'s are integers or half-integers.
On $f(0,b,0,d)$ this imposes the constraint
$$2q_c = (p-q)+(b+d).$$ 
The index of $D_c$ is thus given by the number of ways to 
choose nonnegative integers $b$ and $d$ so that 
$$b+d = 2q_c-(p-q)$$ 
with $p-q$ = $2$ or $0$ plus twice the number of ways to 
choose nonnegative integers $b$ and $d$ so that 
$$b+d = 2q_c-1.$$ 
There are $8q_c$ such solutions for positive $q_c$, one such
solution for $q_c=0$, and none for negative $q_c$. All of these 
solutions are self-dual, so the index of $D_c$ is
$$
{\rm Index}(D_c) = \left\{\begin{array}{ll} 8q_c & \qquad q_c>0
\\ 1 & \qquad q_c=0 \\ 0 & \qquad q_c <0 \end{array} \right\}.
$$
Note that this result assumes a positive coefficient of $K_c$. For
a negative coefficient, the computation proceeds exactly 
as above, provided that we make the following exchanges of coordinates,
\begin{eqnarray}
z_1&\leftrightarrow& \bar{z}_1 \nn
z_2&\leftrightarrow& \bar{z}_2
\end{eqnarray}
This maps $K_c$ to $-K_c$, and flips the sign of $q_c$ in the 
charge constraint above. In other words, the sign condition in the index
formula is really on $a_cq_c$ for each $j$. Thus the index is
$$
{\rm Index}(D_c) = \left\{\begin{array}{ll} 8|q_c| & \qquad a_cq_c>0
\\ 1 & \qquad a_cq_c=0 \\ 0 & \qquad a_cq_c <0 \end{array} \right\}.
$$
for each $c$.

Thus, whenever there exist a massgap, the index ${\cal I}_s^+$ is
$$
{\cal I}_s^+ = \sum\left(\prod_c\left\{\begin{array}{ll} 
8|q_c| & \qquad a_cq_c>0
\\ 1 & \qquad a_cq_c=0 \\ 0 & \qquad a_cq_c <0 \end{array} \right\}\right).
$$
where the sum is over the zeros of the potential. Note that the index 
is nonvanishing only if all $a_cq_c$ (no summation) are nonnegative.
The states in the kernel of the Dirac operator must be annihilated by
${\cal H}-{\cal Z}$ as well, and the central charge ${\cal Z}$ of the states
\begin{equation}
e \sum_c a_cq_c >0
\end{equation}
equals the energy.

Computation of ${\cal I}_s^-$, appropriate for those states with positive 
central charge, proceeds similarly. In fact, this problem can be mapped 
to that of ${\cal I}_s^+$ by
\begin{eqnarray}
\varphi & \rightarrow & \varphi^* \nn
\varphi^* & \rightarrow & \varphi \nn
K_c & \rightarrow & -K_c
\end{eqnarray}
The net effect is to flip the sign condition on the charges $q_c$,
so
$$
{\cal I}_s^- = \sum\left(\prod_c\left\{\begin{array}{ll} 
8|q_c| & \qquad a_cq_c <0
\\ 1 & \qquad a_cq_c=0 \\ 0 & \qquad a_cq_c >0 \end{array} \right\}\right).
$$
whenever a massgap exists. The sum is over zeros of the potential.
The energy of the contributing states is $-e\sum_c a_cq_c >0$.

We consider next the same set of operators but now restricted to the 
$+1$ eigenspace of $\tau_4$. We see that the terms with $p+q$ 
even are in the $+1$ eigenspace of $\tau_4$, and the terms with 
$p+q$ odd are in the $-1$ eigenspace of $\tau_4$. This leads to 
zero index for all nonzero $q_c$. When $q_c=0$, we get a
solution with $p-q=0$. There is only one of these. The index of
$\tau_4$ is then
$${\cal I}_4 = \sum\left(
\prod_c\left\{\begin{array}{ll} 1 & \qquad q_c=0 \\ 0 & \qquad q_c \neq
0 \end{array} \right\}\right)$$
where the sum is over the zeros of the potential.

Finally, we consider the minor modifications necessary to compute
${\cal I}_2$.  Once again a separation of variables allows us to reduce the 
index of the Euclidean operator to a product of indices of operators 
$B_c$, $l = 1,\cdots, k$ on $C^2$. In coordinates, $B_c$ has the form 
$$B_c = \sum_{j=1}^2[\lambda_{2j-1}(\frac{\partial}
{\partial x_j} + iy_j) + \lambda_{2j}
(\frac{\partial}{\partial y_j} - ix_j)],$$
acting on the $+1$ eigenspace of $4\lambda_1\lambda_2\lambda_3\lambda_4$. 
This choice of the Dirac operator corresponds to positive coefficients
$a_c=2$ for all $c=1,\dots,k$. Then in a covariant constant frame, 
$$2B_cB_c^\dagger+2B_c^\dagger B_c = \Delta + |z|^2 + 4iK 
- 4i\sum_j\lambda_{2j-1}\lambda_{2j}.$$
Acting on the algebraic span of $f(a,b,c,d)$,    
$$2B_cB_c^\dagger+2B_c^\dagger B_c = 4(a+c) 
+ 4 + 4i\sum_j\lambda_{2j-1}\lambda_{2j}.$$ 
This has infinite dimensional kernel spanned by the product of 
$f(0,b,0,d)$ 
and a covariant constant spinor in the (one dimensional) 
intersection of the $-1/2$ eigenspaces of 
$i\lambda_1\lambda_2$ and $i\lambda_3\lambda_4.$

The charge constraint in a covariant constant frame takes the form  
$$2q_c = 2K_c/i - i\lambda_1\lambda_2 - i\lambda_3\lambda_4.$$
Acting on the above basis elements of the kernel of 
$2B_cB_c^\dagger+2B_c^\dagger B_c$ 
this reduces to
$$2q_c = b+d + 1.$$
Counting as before this yields a $2q_c$ dimensional kernel 
which lies entirely in the $+1$ eigenspace of 
$4\lambda_1\lambda_2\lambda_3\lambda_4$. Hence,  the index of $B_c$ is
$$
{\rm Index}(B_c) = \left\{ \begin{array}{ll} 2|q_c| & \qquad a_cq_c > 0
\\ 0 & \qquad a_cq_c \le 0 \end{array} \right\}.
$$
Thus, whenever there exist a massgap, the index ${\cal I}_2$ is
$$
{\cal I}_2 = \sum\left(\prod_c\left\{\begin{array}{ll} 
2|q_c| & \qquad a_cq_c > 0
 \\ 0 & \qquad a_cq_c \le 0 \end{array} \right\}\right).
$$

Finally, we conclude the index computations by noting that they
reproduce the four-dimensional results summarized in the previous
section. In fact, the wavefunctions found in Ref.~\cite{1/4} and 
in Ref.~\cite{pope} can be seen easily to reduce to the superharmonic 
oscillator wavefunctions above in the limit of $\tilde a\gg q$.

\section{BPS Bound States}

The above index computations count differences in the number of
ground states with respect to $Z_2$ involutions $\tau_\pm,\tau_4,\tau_2$, 
\begin{equation}
{\rm Index}=n_+ - n_-
\end{equation}
where $n_\pm$ are the number of ground states with $\tau$ eigenvalue
$\pm 1$. We are actually interested in the sum $n_+ + n_-$ instead, 
for which one needs a more refined understanding of the dynamics. For
the case of $\tau_2$ and $\tau_\pm$, we anticipate $n_-$ vanishes
by itself. Such a vanishing theorem is shown rigorously for the 
simplest cases in Appendix. We will assume in this section
that $n_-=0$ holds true for  $\tau_2$ and $\tau_\pm$ in all 
cases, and compare the results to what are expected on physical grounds.

\subsection{$N=4$ Yang-Mills Theories}

The supersymmetric quantum mechanics with four complex supercharges
describe dynamics of monopoles in $N=4$ Yang-Mills theories. Recent
studies of D-branes indicates the following three possibilities for
dyonic bound states of monopoles.
\begin{itemize}
\item 
The state is 1/2 BPS in the Yang-Mills field theory. These states
would be annihilated by all supercharges of the low energy monopole
dynamics, which is possible only if the central charges in the 
relative part of the dynamics is absent. This is guaranteed when
all relative electric charge $q_a$'s vanish. In particular, this
includes purely magnetic bound states.

\item
The state is 1/4 BPS in the Yang-Mills field theory. These states
would be annihilated by half of the supercharges of the low energy
monopole dynamics and not by the other half. This is possible 
only if at least one central charge is nonzero.

\item
The state is non-BPS. 

\end{itemize}
The index computation of the previous section tells us something about
1/2 BPS and 1/4 BPS states, where we counted indices ${\cal I}_4$ and 
${\cal I}_s^\pm$ in the special limit where only one $G^I$, say $G^5$, is 
turned on. Equivalently, we considered vacua where two Higgs fields are 
turned on. 

Of the three indices, only
${\cal I}_4$ is robust against turning on more than one $G^I$'s. 
The Dirac operator $iQ\pm Q^\dagger$ would no longer anticommute with 
$\tau_\pm$ but does anticommutes with $\tau_4$. Only ${\cal I}_4$ is 
a well-defined index in such generic vacua. Turning on additional
$G^I$ always increases the massgap, and is a Fredholm deformation that
preserves ${\cal I}_4$. Thus our result shows that, in generic vacua,
\begin{equation}
{\cal I}_4 = 1 ,
\end{equation} 
when $q_a\equiv 0$, and zero otherwise. Since the central charge of the
state that contributes to the index is zero, the state must be annihilated 
by all supercharges of the quantum mechanics and is a 
1/2 BPS in $N=4$ Yang-Mills theory.\footnote{One might 
think that existence of this bound state is obvious since the potentials are
all attractive and also there exists a classical BPS monopole of the same
magnetic charge. However, none of these guarantee the existence of BPS
bound state at quantum level. In fact, the same set of facts are true 
for a pair of distinct monopoles in $N=2$ $SU(3)$ Yang-Mills theory but
we know that such a purely magnetic bound state does not exist as a
BPS state \cite{n2}.}
This is consistent with the existence of a unique magnetic 1/2 BPS bound state 
of monopoles in generic Coulomb vacua, which is expected from the $SL(2,Z)$ 
electromagnetic duality. One of the generators of $SL(2,Z)$ maps massive 
charged vector multiplets to purely magnetic bound states in 1-1 fashion.
After taking into account the
automatic degeneracy $16$ from the free center-of-mass fermions, the total
degeneracy of these bound states is alway 16, which fits the $N=4$ vector
multiplet nicely. This purely magnetic bound state was previously constructed
by Gibbons \cite{gibbons} in special vacua where all $G^I$'s 
vanishes.

Existence of 1/4 BPS states are more sensitive to the vacuum choice and
the electric charges. The existence criteria were first found by Bergman
\cite{bergman}, where he constructed these dyons as string webs ending
on D3-branes. The first necessary condition is that the string web should
be planar, which is equivalent to the condition that, effectively, 
only one linearly independent $G^I$ is present. This allows us to assume 
without loss of generality that only $G^5$ is turned on, as far as counting
1/4 BPS states are concerned. Thus, the computation of ${\cal I}_s^\pm$
in the previous section is directly applicable.

Secondly, at each junction of the string web, the string tensions must balance 
against each other, which in the present language of low energy dynamics
translates to the condition 
that the effective potential in the charge-eigensector is nonrepulsive along
all asymptotic directions \cite{ly};
\begin{equation}
|q_c|\le |\tilde a_c|.
\end{equation} 
This second condition may indicate the existence of a minimal energy bound 
state, however, does not guarantee that the state would preserve some 
supersymmetry.

Finally, a minimal energy configuration is supersymmetric when the orientation 
of string segments are consistent with each other. Say, if one 
fundamental string segment is directed to one particular direction, 
then another fundamental string in the same web must be directed the same way. 
The second string can point toward the opposite direction and still balance 
the tension, but such a combination breaks all supersymmetry. This 
orientation condition on the string web, is nothing but the condition 
that the product $a_cq_c$'s (no summation) are all of same sign. See
figure 1. Thus, a 1/4 BPS dyon may exist only when $|q_c|\le |\tilde a_c|$
for all $c$ and $a_cq_c$ are all of same sign, at least one of which is 
nonzero. 

\begin{figure}[htb]
\begin{center}
\epsfxsize=5in\leavevmode\epsfbox{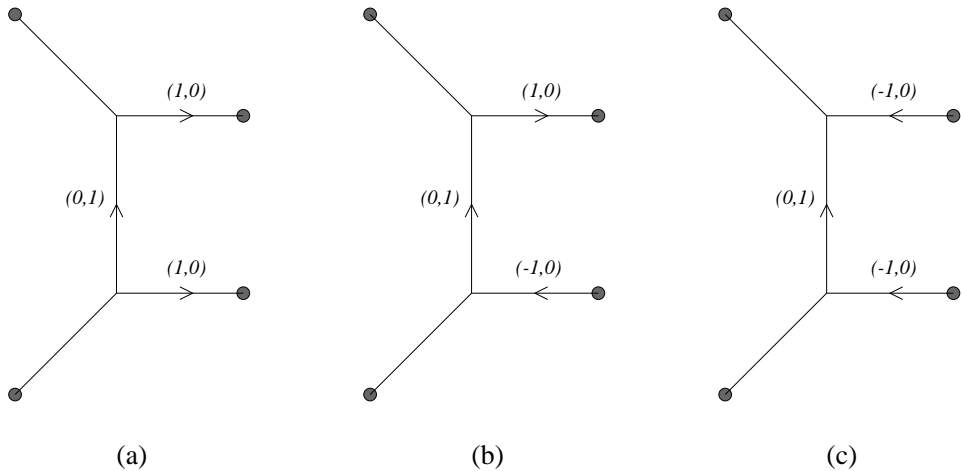}
\end{center}
\caption{Simple webs of $(q,p)$ strings that 
represent dyons in $N=4$ $SU(4)$ theory.
The filled circles represent D3-branes, while vertical lines are D-strings,
and horizontal lines are fundamental strings. Configurations (a) and (c)
preserve a quarter of supersymmetry that was left unbroken by D3's, while 
(b) breaks all supersymmetry. When translated to Yang-Mills field theory
on D3's, the horizontal separations between D3-branes are encoded in $a_c$'s
while the electromagnetic charges are determined by which string ends on which
D3.}
\label{fig1}
\end{figure}

The indices ${\cal I}_s^\pm$ were computed with the massgap condition
$|q_c| < |\tilde a_c|$ to begin with, and yielded nonzero value only when
all $a_cq_c$ were of the same sign; For positive signs of $a_cq_c$, 
${\cal I}_s^+\neq 0$, while for negative $a_cq_c$'s, we have
${\cal I}_s^-\neq 0$. The result is clearly consistent with the existence 
criteria set by the string-web construction, and furthermore gives us 
extra information beyond the string web picture. The index indicates that the 
degeneracy of such a 1/4 BPS state is
\begin{equation}
16\times \prod_c {\rm Max}\,\{8|q_c|,1\}.
\end{equation}
The factor 16 arises from the free center-of-mass fermions. 

In the two monopole bound states, the number $8|q|$ is accounted for by
four angular momentum multiplets of $j=|q|, |q|-1/2, |q|-1/2, 
|q|-1$ \cite{1/4}.\footnote{The first three suffices for $|q|=1/2$.} The top
angular momentum $|q|$ in the relative part of the wavefunction has
a well-known classical origin: when an electrically charged particle moves
around a magnetic object, the conserved angular momentum is shifted
by a factor of $eg/4\pi$. While fermions can and do contribute, the number
of fermions scales with the number of monopoles, and not with the charge $q_a$.
In fact, it is most likely that the top angular momentum of such a dyonic 
bound state wavefunction is
\begin{equation}
j_{top}=\sum_c |q_c|,
\end{equation}
for large charges, so that the highest spin of the dyon
would be
\begin{equation}
1+j_{top}=1+\sum_c |q_c|,
\end{equation}
after taking into account the universal vector multiplet structure 
from the free center-of-mass part. On the other hand, a 1/4 BPS 
supermultiplet with the highest spin $j_{top}+1$ has the total
degeneracy of \cite{ly}
\begin{equation}
16\times 8\sum_c |q_c|,
\end{equation}
which is much less than the number of states we found above unless
all but one $q_c$ vanishes. Thus, this implies that there are many
1/4 BPS, thus degenerate, supermultiplets of dyons for a given set of 
electromagnetic charges. This is probably the least understood of our results.
While one would expect to find degenerate states within a supermultiplet,
there is no natural symmetry that accounts for the existence of
many supermultiplets of the same electromagnetic charges and of the
same energy.

For large electric charges $q_a$, the number of dyon supermultiplets 
scales as, at least,
\begin{equation}
\left(\prod_c {\rm Max}\,\{8|q_c|,1\}\right)/\left(8\sum_c |q_c|\right).
\end{equation}
Proliferation of dyonic states of a given charge was anticipated
by Kol some time ago in the context of string webs in 5 dimensions \cite{kol}. 
Because
our computation was performed for a collection of distinct monopoles, which
put some constraint on the possible topology of the related string web, it
is not immediately clear to us whether we can make any sensible statement
in the regime where Kol's prediction is applicable.\footnote{Kol 
anticipated exponentially large
numbers of states, in fact, which is much more than our powerlike result.}
Nevertheless, it is tantalizing that we found the number of states increasing 
much faster than would have been expected from supersymmetry alone.
It is not clear to us why this happens and what interpretation this
may have in the Yang-Mills field theory.

In the regime where $|q_c| \ge |\tilde a_c|$ for some $q_c$, we cannot rely 
on the current index computation. On the other hand, since even a single 
repulsive direction, i.e., $|q_c| > |\tilde a_c|$ for some $c$, prohibits 
a bound state (supersymmetric or not), the unresolved question boils down 
to the marginal case, where $|q_c|$ equals $|\tilde a_c|$ for some $c$'s 
while the others satisfy $|q_c| < |\tilde a_c|$. The only state that must 
exist for sure is the purely magnetic bound state ($q_a=0$), which was 
constructed by Gibbons when $a_c\equiv 0$ and which is necessary for 
$SL(2,Z)$ invariance. The explicit construction of two-monopole bound states 
in Ref.~\cite{1/4} seem to indicate that no {\it dyonic} bound state may form
along such marginal directions, but this remains to be seen for multi-monopole
cases.

\subsection{$N=2$ Yang-Mills Theories}

In $N=2$ Yang-Mills theories, a state could be either BPS or non-BPS. There
is no such thing as a 1/4 BPS state. Dyons that would have been
1/4 BPS when embedded in $N=4$ theories, are realized as either
1/2 BPS or non-BPS depending on the sign of the electric charges. 
The index of this Dirac operator was nonzero only when
\begin{eqnarray}
0<q_c<\tilde a_c &&\hbox{for all $c$}\nn
\hbox{or}\quad 0>q_c>\tilde a_c &&\hbox{for all $c$}
\end{eqnarray}
which gives us a possible criterion for BPS dyon to exist. This condition
is similar to the condition for BPS dyons or monopoles to exist in $N=4$ 
Yang-Mills theories but differs in two aspects. The first is that given a 
set of $a_c$'s, all of which are positive (negative), the electric charge 
$q_c$'s must be all positive (negative). 

The second difference from $N=4$ case is that a purely magnetic bound state 
of monopoles does not seem to exist as a BPS state, even though there exists a 
classical BPS solution of such a charge. This feature was noted previously
in Ref.~\cite{n2}, where bound states of a pair of distinct monopoles 
were counted explicitly. In fact, the index indicates that
all relative $q_a$ must be nonvanishing for a BPS state to exist.
Assuming the vanishing theorem, the number of BPS dyonic bound state 
under the above condition is
\begin{equation}
4\times \prod_c 2|q_c|,
\end{equation}
The overall factor
$4$ is from the quantization of the free center-of-mass fermions. 

For large electric charges we again observe the proliferation
of supermultiplets. The top angular momentum and thus the size of the
largest supermultiplet can grow only linearly with $\sum |q_c|$ which 
means that the number of supermultiplets with the same electric charges
scales at least as
\begin{equation}
\left(\prod_c 2|q_c|\right) /\left(2\sum_c |q_c|\right)
\end{equation}
for large $q_c$'s. Again it is not clear to us what implication this has
in the Yang-Mills field theory.

In the regime where $|q_c| \ge |\tilde a_c|$ for some $q_c$, again 
we cannot rely on the current index computation. For the same reason as
in $N=4$ Yang-Mills theory, no bound state can exist if even a single 
repulsive direction ($|q_c| > |\tilde a_c|$ for some $c$) exists,
so the unresolved question boils down again to the marginal case, where 
$|q_c|$ equals $|\tilde a_c|$ for some $c$'s while the others satisfy 
$|q_c| < |\tilde a_c|$. Extrapolating from the explicit construction of 
two-monopole bound states in Ref.~\cite{n2}, we suspect that no bound 
state may form along such marginal directions.

\subsection{Ground States of a Noncommutative Instanton Soliton}

Supersymmetric ground states and excited BPS states of an instanton
soliton in $S^1\times R^{3+1}$ can be counted similarly as above. The
only difference as far as computing the index goes, is that the 
potential has many zeros. For a single instanton in noncommutative
$U(n)$ theory, there are precisely $n$ zeros of the bosonic potential,
and near each of these zeros, the Dirac operator can be deformed to
that of a superharmonic oscillator. One crucial difference in 
interpreting the result in physical terms comes from identification
of the conserved charges. Of $n$ possible conserved $U(1)$ charges,
$n-1$ relative charges are again electric charges. However, the overall
conserved $U(1)$ does not correspond to a gauge symmetry, and comes
from translation of the instanton along $S^1$. This last $U(1)$ charge
is just the Kaluza-Klein momentum along $S^1$.

Of particular interest are the quantum ground state of the instanton
with no $U(1)$ charges excited. In the maximally supersymmetric $U(n)$
Yang-Mills theory, the index tells us that there are $n$ distinct
BPS supermultiplets of ground states. This result was anticipated in 
Ref.~\cite{nci}. With half as much supersymmetry, however,
the index is consistent with no supersymmetric quantum ground state exist
at all. 

Quantum states of instanton soliton in $R^4$ was previously studied in 
the commutative setting \cite{ofer,lambert}. In particular,
the absence of a quantum ground state of instanton soliton in the nonmaximal
supersymmetric Yang-Mills theories, has been observed from string-web
construction.

\section{Conclusion}

By computing indices and assuming vanishing theorems, we counted
supersymmetric bound states of arbitrarily many distinct 
monopoles in $N=2$ pure Yang-Mills theories
and also in $N=4$ Yang-Mills theories. The relevant low energy dynamics
are supersymmetric sigma-models with potential(s), where the supercharges
preserved by supersymmetric bound states can be interpreted 
as Dirac operators twisted by triholomorphic Killing vector
fields. An obvious generalization of this computation is to include 
hypermultiplets in $N=2$ Yang-Mills theories, but it goes beyond the
scope of this paper.

Counting of 1/2 BPS states in $N=4$ Yang-Mills yielded a result consistent 
with electromagnetic duality of the theory. In particular, the necessary 
purely magnetic bound states of distinct monopoles are all accounted for 
in $SU(n)$ theories. While this result is not surprising, it is still
significant in that this was shown for the first time in all generic 
Coulomb vacua of the Yang-Mills theory. In contrast, distinct $N=2$ 
monopoles do not seem to bind at all unless all possible relative charges 
are turned on. 

Existence criteria for 1/4 BPS states, previously found in the context 
of string-webs, are also faithfully reflected in the index
formulae. On the other hand, the degeneracy of most 1/4 BPS dyons is 
shown to be much larger that one would have expected from a single 1/4 BPS
supermultiplet with a physically reasonable angular momentum. $N=2$ Dyons
of the same electromagnetic charges as 1/4 BPS dyons of $N=4$ theories,
could be BPS or non-BPS, depending on the signs of the electric charge. 
We also counted  the degeneracy of such $N=2$ BPS dyons,
which shows similar proliferation of supermultiplets. This phenomenon
is not understood at the moment. It should be also interesting to see how the
degeneracy behaves when both electric and magnetic charges are large.

\vskip 1cm
We are grateful to Jerome Gauntlett for stimulating discussions.
We thank Aspen Center for Physics and also the organizers of the workshop,
"The Geometry and Physics of Monopoles", where this work was initiated. 
The work of M.S. is supported by NSF grant DMS-9870161.

\section*{Appendix}

Here, we prove a vanishing theorem for $\tau_2$ and $\tau_\pm$ on
four-dimensional moduli space. Let us consider $\tau_2$ first.
Because of the triholomorphic Killing conditions on $G$, $dG$ 
is self-dual and does not couple to antichiral spinors. Then the 
Dirac operator is a simple Laplacian;
\begin{equation}
-D_\bmu D^\bmu
\end{equation}
when acting on antichiral spinors.
Using the standard trick of sandwiching this operator by a hypothetical
zero mode $\Psi$ and its complex conjugate $\Psi^*$, we find
\begin{equation}
0=-\int \prod_mdz^m\;\Psi^* \nabla_\bmu D^\bmu \Psi=-\int 
\prod_mdz^m\;g^{mn}(D _n \Psi)^*(D_m \Psi)
\end{equation}
where the possible boundary term vanishes by itself since the massgap
forces $\Psi$ to be exponentially small at large distances. Therefore,
\begin{equation}
0=D_m \Psi=(\nabla_m - i G_\bmu)\Psi
\end{equation}
everywhere. This modified connection is still unitary as $G_m$ is real. 
Hence, $\Psi$ is covariant constant with respect to metric compatible
connection and is therefore of constant norm. Such an $f$ cannot be
normalizable on an infinite volume space unless it is 
identically zero, which proves the vanishing theorem in four
dimensions for $\tau_2$.

The  case of $\tau_\pm$ can be handled similarly. Let us recall that
differential forms can be thought of as a tensor product of two spinors.
With an appropriate sign convention, we can identify various sectors of the 
former with those of the latter as follows
\begin{itemize}
\item
selfdual even form $\rightarrow$ [chiral spinor]$\otimes$[chiral spinor]
\item
selfdual odd form $\rightarrow$ [chiral spinor]$\otimes$[antichiral spinor]
\item
antiselfdual odd form $\rightarrow$ 
[antichiral spinor]$\otimes$[chiral spinor]
\item
antiselfdual even form $\rightarrow$ 
[antichiral spinor]$\otimes$[antichiral spinor]
\end{itemize}
On antiselfdual even form, then, the Clifford action of $dG$ is trivial.
Since the self-dual curvature does not couple to antiselfdual forms
either, its action is also trivial. Thus, the square of the Dirac
operator becomes a simple Laplacian again,
\begin{equation}
D_\pm^2=-D_m D^m
\end{equation}
with $D_m=\nabla_m\mp iG^5_m$. By the same logic as in the spinor case, 
therefore, no antiselfdual even-form solution can exist. Finally this 
also shows that antiselfdual odd-form solution does not exist; Unless
the central charge vanishes, a solution generates other solutions via
the action of broken supercharges. The broken supercharges are the
linear combination of $Q$ and $Q^\dagger$ orthogonal to $D_\pm$, so that
it flips $\tau_4$ while preserving $\tau_\pm$. Thus, the number of odd-form
solutions equals the number of even-form solutions, in each $\tau_\pm$
eigensectors, whenever the central
charge is nonzero. This proves the vanishing theorem for $\tau_\pm$ for
sectors with nonzero $U(1)$ charges $q_c$.

\end{document}